\documentstyle[aps,preprint,prl,epsfig]{revtex}
\tightenlines
\title{ Quark Dynamics on Phase-Space }
\author{A. Bonasera\footnote{bonasera@lns.infn.it}}
\address{
Laboratorio Nazionale del Sud, Istituto Nazionale di Fisica Nucleare, 
Via S. Sofia 44, 95123 Catania, Italy
 }
\begin{document}
\maketitle
\begin{abstract}
We discuss the dynamics of quarks within a Vlasov approach. 
We use an interquark ($qq$) potential
consistent with the indications of Lattice QCD calculations and containing 
a Coulomb term,
a confining part and a spin 
dependent term.  Hadrons masses are shown to arise
from the interplay of these three terms plus the Fermi motion
and the finite masses of the quarks.
  The approach gives a lower
and an upper bound for hadrons. The theoretical predictions are shown to
be in fairly good agreement with the experimental data.
\end{abstract}


{\ \vskip 2\baselineskip

\newpage


Very recently the approval and construction of new machines capable of
 accelerating light and heavy ions at 100 GeV/nucleon and more has stimulated
 the
 interest for the search of a quark gluon plasma\cite
 {wong}.  Methods developed to describe the dynamics of two
 colliding heavy ions at low energies 
\cite{repo} can be very useful in this new area
when modified to take into account the intrinsic structure of the hadrons.
  As an
 example we discuss in this paper an approach based
 on the Vlasov equation (VE).  
This equation can be obtained from 
 the quantum Bogoliubov-Born-Green-Kirkwood-Yvon(BBGKY)
 hierarchy by means of the Wigner transform in the limit
 $\hbar \rightarrow 0$\cite{repo,land1}.  
It can include quantum effects such as
 the Pauli principle and can be easily extended to relativistic
 dynamics \cite{wong,repo}. 
  In this paper we will apply the VE
 to the dynamics of the constituents quarks in hadrons.  The quark masses can
 be quite small as in the case of (u,d,s) quarks, (see table I for the values
 used in this paper), so that a relativistic treatment is
 necessary.  As far as the interaction is concerned we will start from a
 two body potential consistent with Lattice Quantum Chromo Dynamics (LQCD)
calculations\cite{wong,pov},
 which displays both a Coulomb $U(r\rightarrow 0) \propto 1/r$ behaviour and
a confining part $U(r\rightarrow \infty) \rightarrow \infty$.  In particular,
we will use the Richardson's potential which depends on the scale parameter
 $\Lambda$\cite{rich}.
From such a starting point we
 calculate the time evolution of the quarks   
and the  masses of the
 hadrons.  The general experimental features are quite reasonably
 reproduced, even though the radii of light hadrons are underestimated.  
We notice that a similar approach has been proposed in \cite{mosel},
but a different interaction was used.

   We briefly recall some general features of the VE, 
interested  people should look the (not complete) list of
    references for more details on this equation
   \cite{wong,repo,land1}.  The VE gives the time evolution of the one
   body distribution function $f_{q_c}(r,p,t)$ in phase-space:
   
   \begin{equation}
\partial_{t}f_{q_c}+\frac{\overrightarrow{\text{p}}}{E}\cdot \nabla _{r}f_{q_c}
-\nabla
_{\text{r}}U\cdot \nabla _{\text{p}}f_{q_c}=0 
\label{lv}
\end{equation}

where U(r) is the potential (discussed below) which governs the quarks dynamics,
$E=\sqrt{p^2+m^2}$ is the energy and 
$m$ is the quark mass. The underscripts  indicate
that the distribution function
 depends on the flavor (q) and color (c) of the quarks.  Notice that the
kinetic part is properly relativistically treated, while the potential
term is non relativistic.  A relativistic extension of the Richardson's 
potential is discussed in \cite{crat} but will not be included here.  

Numerically the VE equation is solved by writing the one body distribution
function as:

\begin{equation}
\label{efer}
f_{q_c}(r,p,t) = \frac{1}{n_{tp}} \sum_i^N g_r(r-r_i(t)) g_p(p-p_i(t)) 
\end{equation}

where the $g_r$ and $g_p$ are sharply peaked distributions (such
 as delta functions, gaussian or other simple
functions), that we shall treat as delta functions.
 $N=Qn_{tp}$ is the number of such terms, $Q={q}+\bar q$ is
the total number of quarks and antiquarks (for a meson Q=2).  Actually, 
N is much larger than the total quark
number Q, so that we can say that each quark 
is represented by ${n_{TP}}$ terms called test particles(tp).
The rigorous mean field limit
can be obtained for ${n_{TP} \rightarrow \infty}$ where the calculations are
of course numerically
 impossible, even though the numerical results converge rather quickly.
 Inserting eq.(~\ref{efer}) in the Vlasov equation
 gives
the Hamilton equations of motions for the t.p.~\cite{repo}:

\begin{eqnarray}
\dot {\bf r}_i = \frac{{\bf p}_i}{E_{i}} \nonumber \\
\dot {\bf p}_i  = - {{\bf\nabla}_{r_i}} U ,   \label{ripi}
\end{eqnarray}
 
 for $ i=1....N$. The total number of tp used in this work
ranges from 5000 to 50000 with no appreciable change in the results.
The equations of motion (3) 
are solved by using a O($\delta t^4 $) Adams-Bashfort
method ~\cite{koon}.
 
Let us now specify the $q\bar q$-potential.  
In agreement to LQCD calculations\cite{pov,rich}
 we have for mesons($\hbar=1$):
 
 \begin{eqnarray}
 U(r)=\frac{8\pi}{33-2n_f}\Lambda(\Lambda r-\frac{f(\Lambda r)}{\Lambda r})
+\frac{8\pi}{9}
\bar{\alpha_s}\frac{<{\bf \sigma}_q{\bf \sigma}_{\bar q}>}
{m_q m_{\bar q}}\delta({\bf r}) 
 \end{eqnarray}
 
 where 
 
 \begin{eqnarray}
f(t)=1-4 \int{\frac{dq}{q}\frac{e^{-qt}}{[ln(q^2-1)]^2+\pi^2}} 
 \end{eqnarray}

 $n_f$
is the number of quark flavors involved and
the parameter $\Lambda$ has been fixed to reproduce the masses of heavy 
$c \bar c$ and $b \bar b$ systems in \cite{rich}.
In eq.(4) we have added to the Richardson's potential 
the chromomagnetic term, very important
 to explain the masses of
different resonances for light quarks.  
In this work the expectation value of $<{\bf\sigma}_q
{\bf\sigma}_{\bar q}>$ is used depending on the relative spin orientations of
the constituent quarks . For instance for the pion this term is
equal to -3 while for a $\rho$ meson it is equal to +1\cite{pov}.  The
$\delta$ function is approximated to a gaussian i.e. we make the 
replacement
$\delta({\bf r})\rightarrow \frac{1}{(2\pi\sigma^2)^{3/2}}e^{-\frac{r^2}
{2 \sigma^2}}$, and we fixed $\sigma=0.5 fm$. 
 The results depend on the ratio of the average value
of the strong coupling constant $\bar \alpha_s$ to the variance of the
gaussian.  Such ratio was fixed to fit the mass difference between $\pi$
 and $\rho$ mesons (see below).   
  Clearly the chromomagnetic interaction becomes unimportant for
very heavy quarks.  The potential term eq.(4) acts between two quarks, since
we are describing quarks as a swarm of tp, we normalize the potential by
a factor $1/n_{tp}$ \cite{belk}.

   The initial conditions are given
 by  randomly distributing the tp in a sphere of radius r in coordinate
space and $p_f$ in momentum space.  $p_f$ is the Fermi momentum
estimated in a simple Fermi gas model by imposing that a cell in
phase space of size $h=2 \pi$ can accommodate at most two identical quarks
 of different spins. A
simple estimate gives the following relation between the quarks density
$n_q$ and the Fermi momentum:
\begin{eqnarray}
n_q=\frac{g_q}{6\pi^2}p_f^3 
 \end{eqnarray}
an analogous formula can be derived for $\bar q$\cite{wong}.
  The degeneracy number 
$g_q=n_c\times n_s \times n_f$, where $n_c$ is the number of colors and
$n_s$ is the number of spins\cite{wong}.  For quarks and
 antiquarks 3 different colors are used red,green and blue (r,g,b) \cite{pov}.  
  From the above equation
we see that the Fermi momentum for quarks distributed in a sphere of radius
0.5 fm is of the order of 0.5 GeV/c. Thus relativistic effects become important
for quark masses less than 1 GeV. For instance for the $\pi$ case discussed
below, if we calculate the total energy of the system 
relativistically we obtain 0.14 GeV,
 while using the nonrelativistic limit we obtain about 1 GeV!  
We stress that since the system is properly antisymmetrized at time t=0fm/c, 
it will remain so at all times since the VE conserves the volume in phase-space
\cite{repo}.

The masses of the hadrons are determined by finding the minumim 
total energy of the system as a function of the initial radius r.
For each initial radius r a Fermi momentum is deduced from eq.(6) and
this gives (after adding the potential term) a total energy of the system 
which is of course constant in time.  
Changing the initial radius changes the total energy,
and it has a minimum at $r=\bar r$.  Thus 
it is the interplay among the Fermi motion, 
the potential term and the quark masses 
which determines the masses of the hadrons.  

In order to test our approach, 
we have first 
studied heavy quark systems using the same values for
the $c$ and $b$
 masses, $n_f=3$ and $\Lambda=0.398 GeV$ as in ref.\cite{rich}
and neglecting the chromomagnetic term.  We
obtained the minimum mass values of 2.9 and 9.5 GeV 
for the $c\bar c$ and $b\bar b$ systems, to be contrasted with 3.1 GeV and
9.4 GeV obtained in \cite{rich}.  
The good agreement to
the quantum calculation of \cite{rich} suggests that the VE is quite well
justified and in particular the mean field approximation is good.
This is probably due to the fact that the interparticle potential is the
result of many gluons exchanges and that the hadrons are made of a
combination of colored quarks and in this sense the mean field approximation
is reasonable.  Also, it is important to stress that in a quantum calculation
 the bare two body potential is folded with smooth $q\bar q$ wave-functions.
 The smoothness of the wave functions is simulated in the VE 
with the use of (a large number of) tp.

In order to study the limits and merits 
of our approach we have extended the calculations
to lighter quarks.  For such systems the chromomagnetic term is important
to reproduce the experimental values of the masses.  Using the same value
of the scale parameter $\Lambda=0.398 GeV$ as in \cite{rich} and fixing the
(u,d) quark masses and the strong coupling constant $\bar \alpha_s$,
 we can easily reproduce
the pion mass, but not the $\rho$ mass at the same time.  
Thus we have readjusted the values
of the scale constant and the 
quark masses to reproduce the data.  We found a good fit for 
mesons by using $\Lambda=0.250 GeV$,
 $\bar \alpha_s=0.225$ \cite{pov} and
the quark mass values given in Table I.  We notice that
the parameters and the heavy quark mass values 
are in good agreement with  currently accepted ones \cite{pov,data}. The masses
of (u,d) quarks in table I is somewhat smaller than the 300 MeV 
used in many potential models \cite{pov}.  This is due to the fact that these
models are non relativistic but, as we stressed above, because of the Fermi
motion, relativistic effects are quite important.
In the relativistic approach of \cite{crat}, where the Richardson's potential
was used as well, the (u,d,s) quark masses have values comparable to ours. 
The small differences 
between our results and \cite{crat} are most probably due to their
relativistic generalization of the Richardson's potential.

 The parameters entering our model are essentially fixed on some
meson masses.  As a consistency check we extended the calculations
to the baryons case with the usual modifications of a factor $1/2$ 
to the potential eq.(4), as 
suggested by LQCD considerations\cite{pov}.

From the knowledge of the distribution function at each time step, we
 can easily calculate the density and the potential 
of the system at one time step.
An average over time of these quantities can be performed as 
  well.  In the calculations the time averages were performed over a time
 interval up to 100 fm/c. 
In figure (1), we plot the density (left column) 
as a function of the distance
$r$ from the center of the system and obtained
  as average over the one body distribution 
  function at one time step(square symbols), and over time as well (full line). 
The top panel correspond
  to a total energy of a (u,d) system of 0.140 GeV and a root mean square (RMS)
 radius (averaged over time) of
  0.33 fm, i.e. the pion.   We stress that this is the minimum for the total
energy obtained when the total quarks spin add to S=0, i.e. a
 pseudoscalar mesons \cite{pov}.
  The middle panel corresponds to a total energy of
  0.85 GeV and a RMS of .53 fm for a (uud) system,i.e. a nucleon.
A similar calculation but for S=3/2 ($\Delta$)
gives a mass of 1.3 GeV and a RMS of
0.75 fm (bottom panel in figure 1).  In all cases, time and tp averages are
identical which implies that the systems are in equilibrium.
The densities obtained
averaging over the distribution function at one time step, do not extend to
very low values because of the finite number of tp.
The rms radius of the $\pi$ and $n$ are smaller than data \cite{pov},
however the densities extend much further than the rms radius, and  
  they fall off
exponentially similarly to experiments.
 For increasing quark masses the rms
becomes more reasonable as compared to data or other calculations.  

The average potential can be easily estimated as \cite{belk}:

\begin{eqnarray}
\bar U(r)=\frac{1}{n_{tp}}\sum_j U(r,r_j)
\end{eqnarray}

The average potential displays some interesting features, see 
fig.1(right column).  First, all
the potentials go to zero for all systems (even though it is explicitly shown
in the figure for the $\Delta$-case only), 
and for large distances in contrast to the bare
potential eq.(4) that diverges linearly, just because the densities
 go to zero.  The average potential for
$\pi$ and $n$ is an attractive pocket, 
and the confining term gives some contribution at large distances. 
 For the $\Delta$ case,
the chromomagnetic term is repulsive thus the RMS of the system is larger than
 $n$ and the confining term only is responsible for keeping the system bound.
 Looking at the $\Delta$-density we see that the repulsion at small distances
gives a smaller and flatter density as compared to the $n$ case. 

In figure (2) we display the calculated (open symbols)
mass of the resonances vs. the sum of the quark 
masses (cfr. table I),
for mesons (top) and baryons (bottom).  The circle symbols refer to 
attractive, while the squares to repulsive chromomagnetic term in eq.(4)
\cite{pov}. The corresponding 
experimental data \cite{data} are given by the full symbols.
  The overall agreement
is quite good in all cases and some predictions for resonances not yet 
observed are also given.

In order to understand how other higher resonances appear for fixed quark types,
we 
define a dipole operator for mesons,
 analogous to the nuclear case \cite{repo,schuck}:

\begin{eqnarray}
D(t)=\sum_q p_z(t) -\sum_{\bar q}p_z(t)
\end{eqnarray}

where the sum is extended over all the tp for $q$ and $\bar q$, the choice of
the z-axis is of course arbitrary. Resonances are better seen by defining
the dipole strength function S(E), where E is the energy:

\begin{eqnarray}
S(E)=|F(E)|^2
\end{eqnarray}

with the Fourier transform

\begin{eqnarray}
F(E)=\int dt{[e^{i(E t/\hbar)}D(t)]}
\end{eqnarray}

In figure 3 we plot the function S(E) vs E for the $\pi$ (ud-quarks) case.
  Strong resonances 
extend to about 1. GeV.  This implies that we can have mesons up to 1. GeV 
above the ground state mass.  In the case of the heavier $b\bar b$ quarks,
resonances extend to about 0.5 GeV above the ground state.  Experimentally 
resonances are seen for instance for the $b\bar b$ case to about 2 GeV above the
smallest resonance \cite{data}.  In our model
we find that to obtain resonances at higher energies we need to 
increase the strength of the confining term.  Thus the measured upper values 
of the hadron masses can give a constraint on the value of the string tension.

The results discussed above prove that the VE is suitable to
describe the quarks dynamics in hadrons despite the simple form for
the two body potential.  The approach works rather well for heavy quarks, in
that it is able to reproduce the masses and radii of the heavy hadrons
(as compared to data or other calculations \cite{pov,rich}) and also the
 masses of light hadrons.  The radii of light quarks systems are
 underestimated  which could be a hint for relativistic corrections to the 
potential term as discussed for instance in \cite{crat}.

The method is quite easy to implement and the numeric is rather well under
control. The equation of state of quark matter can be calculated within the
same formalism.  
Future work will be also to implement a collision term which should
help to understand the dynamics of colliding hadrons.

 {\ \vskip 0.7 cm \centerline{\bf ACKNOWLEDGMENTS} }

 We acknowledge the Hadron group at JAERI, where this work was started,
 for financial support,
kind hospitality and stimulating discussion.    We also thank
profs. M.Consoli and A. Kerman for useful suggestions and comments.


\newpage

{\ \centerline 
{\bf Table I} }

Quark masses used in this work.

\vspace{0.7cm}

\begin{tabular}{|c|c|}
\hline
$Quark Mass $ &GeV   \\ \hline
$u$ & 0.13  \\ 
$d$ & 0.13  \\ 
$s$ & 0.35   \\ 
$c$ & 1.45   \\
$b$ & 4.8   \\
$t$ & 180.    \\[3mm] \hline
\end{tabular}

\newpage

\begin{figure}[tbp]
\begin{center}
\mbox{{ \epsfysize=14 truecm \epsfbox{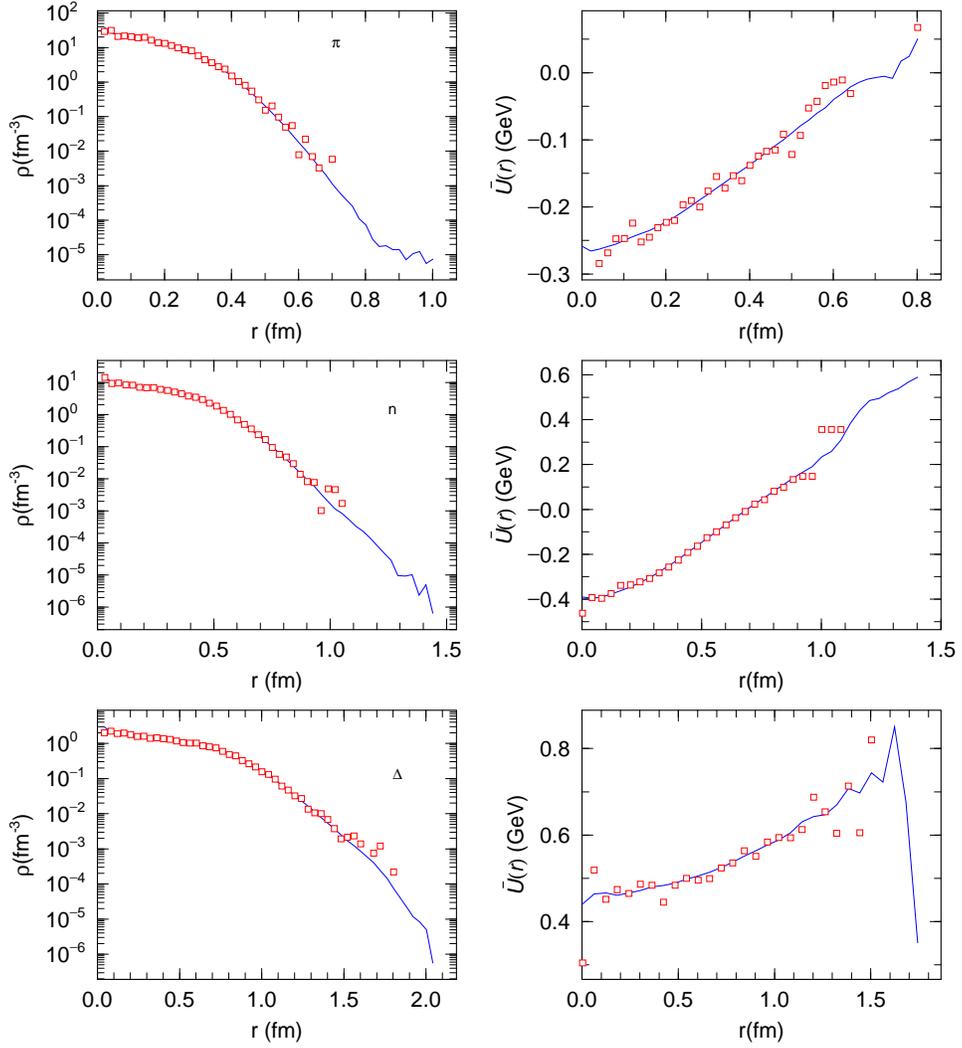}}}
\vskip 1.5cm
\caption{Average density (left column)
for a $\pi$(top), $n$(middle) and a $\Delta$
(bottom) hadrons.  The average is over tp (squares) and
over tp and time (full line). Similarly for the average potential (right 
column).\label{fig:fig1}}
\end{center}
\end{figure}
\begin{figure}[tbp]
\begin{center}
\mbox{{ \epsfysize=14 truecm \epsfbox{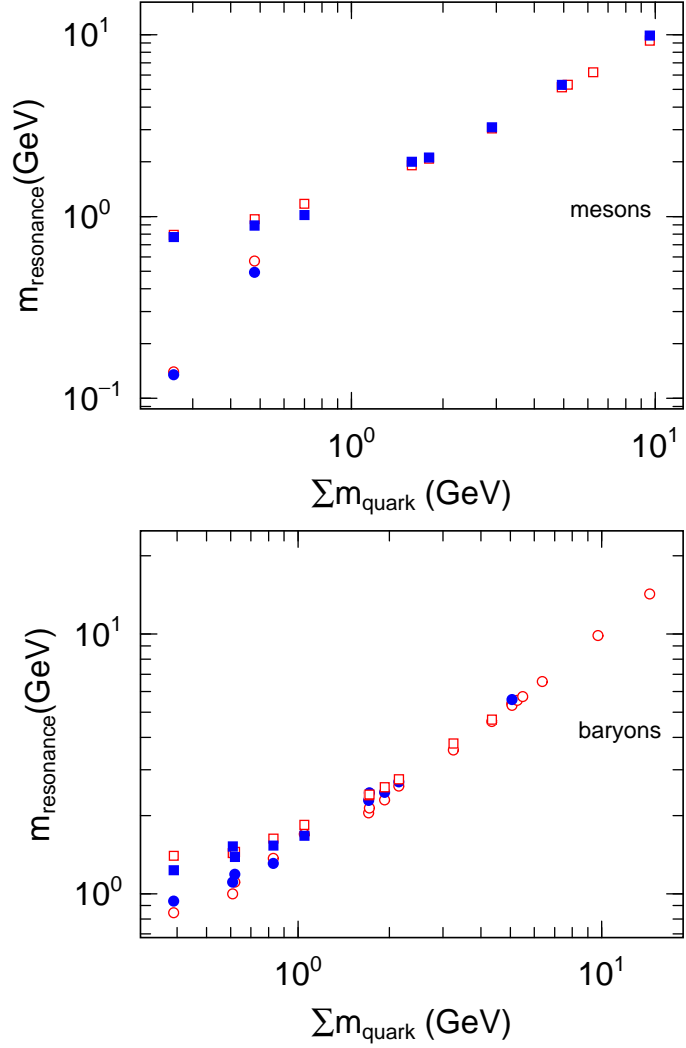}}}
\vskip 1.5cm
\caption{Meson masses (top) vs. masses of the $q\bar q$ pair. The symbols
refer to the observed (full symbols) and calculated (open
symbols) masses. The circles refer to the pseudoscalar and the
 squares to vector mesons masses. 
 Similarly for baryons (bottom). Data are taken from [9].\label{fig:fig2}}
\end{center}
\end{figure}
\begin{figure}[tbp]
\begin{center}
\mbox{{ \epsfysize=14 truecm \epsfbox{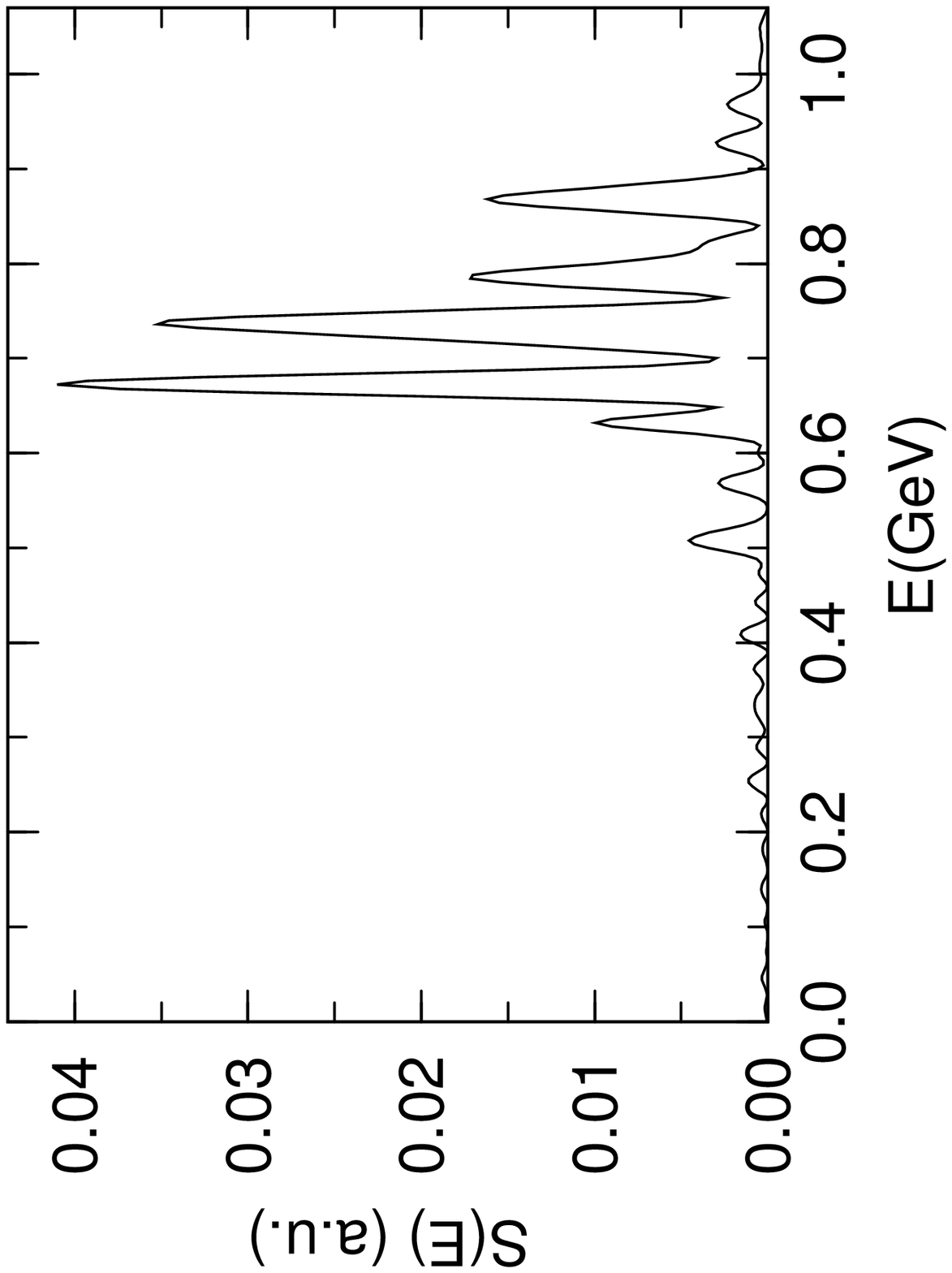}}}
\vskip 1.5cm
\caption{Strength function (in arbitrary units)
 versus energy  for the $\pi$ case.\label{fig:fig3}}
\end{center}
\end{figure}

\newpage 

\begin{references}
 \bibitem{wong} C.Y. Wong, {\it Introduction to High-Energy Heavy ion
  Collisions},
 World Scientific Co., Singapore,1994;  L.P. Csernai {\it Introduction
 to Relativistic Heavy ion
  Collisions}, John Wiley and Sons, New York, 1994.

\bibitem{repo} A. Bonasera, F. Gulminelli and J. Molitoris, Phys. Rep.
{\bf 243}, (1994)1;  G. Bertsch, S. Dasgupta, 
Phys. Rep.{\bf 160}(1988){189}, and references therein.

\bibitem{land1}E.M. Lifshitz and L.P. Pitaevskii,
{\it Physical Kinetics},  Pergamon Press 1991.

\bibitem{pov} B.Povh, K.Rith, C. Scholz, F.Zetsche, {\it Particles and Nuclei:
an introduction to the physical concepts}, Springer, Berlin, 1995.

\bibitem{rich} J.L. Richardson, Phys.Lett. {\bf 82B}(1979)272.

\bibitem{mosel} T.Vetter, T.Biro and U.Mosel, Nucl.Phys.{\bf A581}(1995)598;
S.Loh, T.Biro, U.Mosel and M.Thoma, Phys.Lett.{\bf B387}(1996)685;
S.Loh, C.Greiner, U.Mosel and M.Thoma, Nucl.Phys.{\bf A619}(1997)321;
S.Loh, C.Greiner and U.Mosel, Phys.Lett.{\bf B404}(1997)238.

\bibitem{crat} H.W. Crater, P. Van Alstine, Phys.Rev.Lett.{\bf 53}(1984)1527.

\bibitem{koon} S.E.Koonin, D.C. Meredith, {\it Computational Physics},
Addison-Wesley publ.c.,USA, 1990.

\bibitem{belk} M.Belkacem, V.Latora and A. Bonasera, Phys.Lett.{\bf 326B}
(1994)21.




\bibitem{data}
R.M. Barnett et al. Phys.Rev.{\bf D54},(1996)1.

\bibitem{schuck}
P.Ring, P.Schuck, {\it The Nuclear Many-Body Problem},Springer-Verlag,
New York, 1980.

 \end{references}
\end{document}